\documentclass[%
 aip,
 amsmath,amssymb,
 reprint,%
]{revtex4-1}

\usepackage{braket}
\usepackage{graphicx}% Include Fig. files
\usepackage{dcolumn}% Align table columns on decimal point
\usepackage{bm}% bold math
\usepackage[utf8]{inputenc}
\usepackage[T1]{fontenc}
\usepackage{mathptmx}
\usepackage{etoolbox}

\makeatletter
\def\@email#1#2{%
 \endgroup
 \patchcmd{\titleblock@produce}
  {\frontmatter@RRAPformat}
  {\frontmatter@RRAPformat{\produce@RRAP{*#1\href{mailto:#2}{#2}}}\frontmatter@RRAPformat}
  {}{}
}%
\makeatother
\begin{document}

\preprint{AIP/123-QED}

\title{Nonadiabatic Dynamics Near Metal Surface With Periodic Drivings: \\ A Floquet Surface Hopping Algorithm}
% Force line breaks with \\
\author{Yu Wang}
\email{wangyu19@westlake.edu.cn}
 \affiliation{Department of Chemistry, School of Science, Westlake University, Hangzhou 310024 Zhejiang, China}%Lines break automatically or can be forced with \\
 \affiliation{Institute of Natural Sciences, Westlake Institute for Advanced Study, Hangzhou 310024 Zhejiang, China}
\author{Wenjie Dou}%
 \email{douwenjie@westlake.edu.cn}
 \affiliation{Department of Chemistry, School of Science, Westlake University, Hangzhou 310024 Zhejiang, China}
 \affiliation{Institute of Natural Sciences, Westlake Institute for Advanced Study, Hangzhou 310024 Zhejiang, China}
 \affiliation{Department of Physics, School of Science, Westlake University, Hangzhou 310024 Zhejiang, China}
  %\homepage{http://www.Second.institution.edu/~Charlie.Author.}

\date{\today}% It is always \today, today,

\begin{abstract}
We develop a Floquet surface hopping (FSH) approach to deal with nonadiabatic dynamics of molecules near metal surfaces subjected to time-periodic drivings from strong light-matter interactions. 
The method is based on a Floquet classical master equation (FCME) derived from a Floquet quantum master equation (FQME), followed by a Wigner transformation to treat nuclear motion classically. We then propose different trajectory surface hopping algorithms to solve the FCME. We find that a Floquet averaged surface hopping with electron density (FaSH-density) algorithm works the best as benchmarked with the FQME, capturing both the fast oscillations due to the driving as well as the correct steady state observables. This method will be very useful to study strong light-matter interactions with a manifold of electronic state. 

\end{abstract}

\maketitle

\section{\label{sec:level1}Introduction}
Floquet engineering is a powerful tool to control the dynamics of quantum systems using time-periodic external fields (e.g. electromagnetic fields), which is account for various phenomena\cite{floquet2019}. In practice, surface plasmon or a Fabry-P\'{e}rot cavity can create a resonant electromagnetic (EM) field and realize Floquet engineering through strong light-matter interactions. For instance,  
strong light-matter coupling is achieved by placing materials in a resonance EM field and show a great promise to control properties of matter without structural modifications\cite{csr2019,sci_minireview2021}. In addition, n-type or p-type organic semiconductors deposited on top of a metal surface, which acts as an open electromagnetic cavity, have been observed a noteworthy conductivity enhancement\cite{orgiu2015conductivity,nagarajan2020conductivity}.
Moreover, strong light-matter coupling can largely extend spatial range of energy transport in 
organic materials\cite{gonzalez2015harvesting,hou2020ultralong,zhong2017energy}. Furthermore, 
numerous photophysical and photochemical properties, such as electronic relaxation 
pathways\cite{lidzey2002experimental,coles2011vibrationally,coles2013imaging}, quantum 
yields\cite{wang2014quantum,ballarini2014polariton,grant2016efficient,stranius2018selective},, and 
chemical reactivities\cite{hutchison2012modifying,lin2016photoswitchable,feist2018polaritonic,li2021cavity}, can also be manipulated in the strong coupling regime. The microscopic mechanisms leading 
to these modifications through strong light-matter interactions, however, remain largely unexplained.

On the one hand, without Floquet driving, when nuclei interact non-adiabatically with a manifold electronic states (e.g. an adsorbate at a metal surface), there is a drastic breakdown of the Born-Oppenheimer approximation.
There are several numerically exact solutions for such non-adiabatic processes
when only a few degrees of freedom (DoFs) are considered, including multiconfiguration time dependent Hartree (MCTDH)\cite{PhysRevB.76.153313}, quantum Monte Carlo (QMC)\cite{PhysRevLett.100.176403},numerical renormalization group (NRG)\cite{NRG_1,NGF_2,NRG_3}, and hierarchical quantum master equation (HQME)\cite{HQME_1,HQME_2}. That being said, these methods are difficult to apply to large, realistic systems. Many attempts have been given to develop approximate methods to deal with realistic systems involving nonadiabatic dynamics in open quantum systems\cite{galperin2007molecular,reimers2007green,bode2012current,akimov2013pyxaid}. On the other hand, when considering nonadiabatic dynamics under Flquet driving, there exists Floquet surface hopping\cite{fiedlschuster2016floquet,zhou2020nonadiabatic,chen2020proper}, as well as coupled-trajectory mixed quantum–classical\cite{schiro2021quantum} methods to study dynamics for a closed system. For open quantum systems, there are far less approaches available with only a few exceptions, 
such as Floquet scattering matrix\cite{floquet_scatter_matrix,floquet_scatter2014,floquet_scatter2019}, 
Floquet Green's functions\cite{floquet_green2000,floquet_green,wu2010noise,honeychurch2023quantum}, and Floquet master equations\cite{floquet_QME1997, floquet_QME2010,kuperman2020mechanical}. Again, these methods are only applicable to small systems. Accurate methods are needed to study large and realistic systems. 

In this article, we attempt to offer an accurate and efficient tools to study nonadiabatic dynamics in open quantum systems with Floquet engineering. We first derive a Floquet quantum master equation (FQME)
to accurately uncover the non-adiabatic processes 
of a periodically driven impurity level near a metal surface.
Then a corresponding Floquet classical master equation (FCME) is achieved by taking the Wigner transformation of FQME, similar to the steps in Ref. \cite{SH_2} for non-Floquet scenario. We then propose three Floquet surface hopping algorithms to solve the FCME. We find that an algorithm named time-averged Floquet surface hopping with density (FaSH-density) is capable to reproduce fully dynamics as well as steady state results as benchmarked against FQME. 

An outline of this paper is as follows. In Sec. \uppercase\expandafter{\romannumeral2}, we outline how to derive the FQME and FCME, and present the FSH algorithm. In Sec. \uppercase\expandafter{\romannumeral3}, we present results of both  electronic population and phonon relaxation dynamics under different Floquet drivings employing five different methods (FQME, FSH, FaQME, FaSH, and FaSH-density). We conclude in Sec. \uppercase\expandafter{\romannumeral4}.

\section{Theory}
\subsection{Model Hamiltonian}
We consider the Anderson-Holstein (AH) Hamiltonian with one molecule coupled to a Fermionic bath as well as subjected to periodic driving. The molecule consists of a single electronic level coupled to a single phonon and a manifold of electrons:
\begin{equation}\label{eq0}
    \hat{H}(t) = \hat{H}_S(t) + \hat{H}_B + \hat{H}_T
\end{equation}
\begin{equation}\label{eq1}
\begin{split}
    \hat{H}_S(t) = (E_d + A\sin(\Omega t))d^+d +
    & g(a^+ + a)d^+d + \\& \hbar\omega(a^+a + \frac{1}{2})
\end{split}
\end{equation}
\begin{equation}\label{eq2}
    \hat{H}_B = \sum_k \epsilon_k c_k^+ c_k
\end{equation}
\begin{equation}\label{eq3}
    \hat{H}_T = \sum_k V_k(d^+ c_k + c_k^+ d)
\end{equation}
$\hat{H}_S(t)$ is the system Hamiltonian with $E_{d}$ being the energy level and $\omega$ being the frequency of the harmonic oscillator. $g$ represents the electron-phonon (el-ph) coupling strength.
$A$ is the driving amplitude and $\Omega$ is driving frequency. In addition, 
$\hat{H}_B$ represents the bath Hamiltonian with $\epsilon_k$ being the energy level of the Fermion $c_k$ in the bath. $\hat{H}_T$ is the interaction Hamiltonian with $V_k$ being the coupling between the impurity level $d$ and the Fermion $c_k$ in the bath.

It will be convenient to replace $a^+$ and $a$ with the position $x$ and momentum $p$ (i.e. $x = \sqrt{\frac{\hbar}{2m\omega}}(a^+ + a)$ and $p = i\sqrt{\frac{m\hbar\omega}{2}}(a^+ - a)$), where $m$ is the nuclear mass, such that the system Hamiltonian can be written as
\begin{equation}\label{eq4}
    \hat{H}_S = (E_d + A\sin(\Omega t))d^+d + \sqrt{\frac{2m\omega}{\hbar}} gxd^+d + \frac{p^2}{2m} + \frac{1}{2}m\omega^2x^2
\end{equation}
To derive a Floquet quantum master equation for treating the dynamics of the periodically driven open quantum system, we first separate the time independent part from the time dependent part in $H_S(t)$, such that
\begin{equation}\label{eq5}
    \hat{H}_S(t) = \hat{H}_{mol} + A\sin(\Omega t)d^+d
\end{equation}
\begin{equation}\label{eq6}
    \hat{H}_{mol} = (E_d + \sqrt{\frac{2m\omega}{\hbar}}gx)d^+d + \frac{p^2}{2m} + \frac{1}{2}m\omega^2x^2
\end{equation}
We can further write $H_{mol}$ as
\begin{equation}
    \hat{H}_{mol} = H_0\ket{0}\bra{0} + H_1\ket{1}\bra{1}
\end{equation}
\begin{equation}
    \hat{H}_0 = \frac{p^2}{2m} + \frac{1}{2}m\omega^2x^2
\end{equation}
\begin{equation}
    \hat{H}_1 = \frac{p^2}{2m} + \frac{1}{2}m\omega^2x^2+ \sqrt{\frac{2m\omega}{\hbar}}gx + E_d
\end{equation}
where $\ket{0}$ ($\ket{1}$) denotes unoccupied (occupied) state of the impurity. 

\subsection{Floquet Quantum Master Equation}

We now sketch out the derivation of a Floquet Quantum master equation in treating the dynamics of the system. The key quantity of interest is the reduced density matrix of the system.
Starting with the quantum Liouville equation in the interaction picture,
we find the total density matrix can be expressed as
\begin{equation}\label{eq7}
    \frac{d\rho(t)}{dt} = -\frac{i}{\hbar}[H_T(t),\rho(0)] - \frac{1}{\hbar^2}\int_0^tdt'[H_T(t),[H_T(t'),\rho(t')]]
\end{equation}
where
\begin{equation}\label{eq8}
    H_T(t) = U^+(t)H_TU(t)
\end{equation}

%%%%%%%%%%%%%%%%%%%
Here, the time evolution operator $U(t)$ is
\begin{equation}\label{eq9}
\begin{split}
    U(t) & = \hat{T}\exp\left[-\frac{i}{\hbar}\int_0^t dt'(H_B+H_S(t'))\right] \\
    & = \hat{T}\exp\left[-\frac{i}{\hbar}\int_0^t dt'(H_B+H_{mol}+A\sin(\Omega t')d^+d)\right] \\
    & = \exp(-i(H_B+H_{mol})t/\hbar-ig(t)d^+d/\hbar)
\end{split}
\end{equation}
where $\hat{T}$ is the time ordering operator and $g(t)=\frac{A}{\hbar\Omega}(1-\cos(\Omega t))$

In the Born-Markovian approximation, 
we replace $\rho(t')$ in Eq. (\ref{eq7}) by $\rho_B^{eq}\bigotimes\rho_S(t)$ 
in the integrand that relies on the assumptions that the bath remains in equilibrium 
throughout the process and that bath correlation functions decay fast on the system time scale. 
This leads to (setting $\tau = t-t'$):
\begin{equation}\label{eq16}
\begin{split}
    \frac{d\rho(t)}{dt} = & -\frac{i}{\hbar}[H_T(t),\rho(0)] - \\ &
    \frac{1}{\hbar^2}\int_0^\infty d\tau[H_T(t),[H_T(t-\tau),\rho_B^{eq}\bigotimes\rho_S(t)]]
\end{split}
\end{equation}
Next, we assume the initial total density matrix is a direct product of the system density matrix and the equilibrium bath density matrix, i.e., $\rho(0) = \rho_B^{eq}\bigotimes\rho_S(0)$ and take the trace of Eq. (\ref{eq7}) over bath degrees of freedom. We also use $Tr_B(H_T(t)\rho_B^{eq}) = 0$, which yields
\begin{equation}\label{eq17}
\begin{split}
    \frac{d\rho_S(t)}{dt} = - \frac{1}{\hbar^2}\int_0^\infty d\tau Tr_B[H_T(t),[H_T(t-\tau),\rho_B^{eq}\bigotimes\rho_S(t)]]
\end{split}
\end{equation}

Back to Schr$\rm\ddot{o}$dinger picture, 
$\rho_S(t) = e^{iH_{mol}t}\rho_se^{-iH_{mol}t}$, 
and assuming $\rho_s = \rho_0\ket{0}\bra{0}+\rho_1\ket{1}\bra{1}$, 
which ensures that there will be no coherence between occupied and unoccupied states at later time. 
The reduced density matrix for state 0 (unocupied) and state 1 (occupied) evolves as:
\begin{equation}\label{eq35}
\begin{split}
    \frac{d\rho_0}{dt} = & -i[H_0,\rho_0]-\sum_k\frac{|V_k|^2}{\hbar^2} \times \\ &
    \int_0^\infty d\tau [ e^{i\epsilon_k\tau/\hbar-i(g(t)-g(t-\tau))/\hbar}f(\epsilon_k) e^{-iH_1\tau/\hbar}e^{iH_0\tau/\hbar}\rho_0 \\ & - e^{i\epsilon_k\tau/\hbar-i(g(t)-g(t-\tau))/\hbar}(1-f(\epsilon_k)) \rho_1 e^{-iH_1\tau/\hbar}e^{iH_0\tau/\hbar} \\ &
    + e^{-i\epsilon_k\tau/\hbar+i(g(t)-g(t-\tau))/\hbar}f(\epsilon_k) \rho_0 e^{-iH_0\tau/\hbar}e^{iH_1\tau/\hbar} \\ &
    - e^{-i\epsilon_k\tau/\hbar+i(g(t)-g(t-\tau))/\hbar}(1-f(\epsilon_k))  e^{-iH_0\tau/\hbar}e^{iH_1\tau/\hbar} \rho_1 ]
\end{split}
\end{equation}
\begin{equation}\label{eq36}
\begin{split}
    \frac{d\rho_1}{dt} = & -i[H_1,\rho_1]-\sum_k\frac{|V_k|^2}{\hbar^2} \times \\ &
     \int_0^\infty d\tau [ e^{-i\epsilon_k\tau/\hbar+i(g(t)-g(t-\tau))/\hbar}(1-f(\epsilon_k)) e^{-iH_0\tau/\hbar}e^{iH_1\tau/\hbar}\rho_1 \\ &
    - e^{-i\epsilon_k\tau/\hbar+i(g(t)-g(t-\tau))/\hbar}f(\epsilon_k) \rho_0 e^{-iH_0\tau/\hbar}e^{iH_1\tau/\hbar} \\ &
    + e^{i\epsilon_k\tau/\hbar-i(g(t)-g(t-\tau))/\hbar}(1-f(\epsilon_k)) \rho_1 e^{-iH_1\tau/\hbar}e^{iH_0\tau/\hbar} \\ &
    - e^{i\epsilon_k\tau/\hbar-i(g(t)-g(t-\tau))/\hbar} f(\epsilon_k)  e^{-iH_1\tau/\hbar}e^{iH_0\tau/\hbar} \rho_0 ]
\end{split}
\end{equation}
Here, $f$ is the Fermi function ($f(x)=1/(1+e^{\beta x})$). By employing the Jacobi-Anger expansion, we can express $e^{ig(t)}$ as:
\begin{equation}
\begin{split}
    e^{ig(t)} = e^{\frac{i}{\hbar}\frac{A}{\Omega}}e^{\frac{i}{\hbar}\frac{A}{\Omega}\cos{(\Omega t)}} = e^{\frac{i}{\hbar}\frac{A}{\Omega}}\sum_{n=-\infty}^{+\infty}(i)^n J_n(z)e^{in\Omega t}
\end{split}
\end{equation}
where $n$ is the integer, $J_n(z)$ is the $n$-th Bessel function of the first kind, and $z=\frac{A}{\hbar\Omega}$. Thus, we can expand the term $e^{-i(g(t)-g(t-\tau))/\hbar}$ appears in Eqs. \ref{eq35} and \ref{eq36} as:
\begin{equation}\label{eq37}
\begin{split}
    e^{-i(g(t)-g(t-\tau))/\hbar} = \sum_{n,m}(i)^n(-i)^m J_n(z) J_m(z) e^{i(n-m)\Omega t}e^{im\Omega\tau}
\end{split}
\end{equation}

%If we take the time cyclic average, Eq. (\ref{eq37}) then becomes
%\begin{equation}\label{eq38}
%\begin{split}
 %   e^{-i(g(t)-g(t-\tau))/\hbar} = \sum_{n} J_n(z)^2  e^{im\Omega\tau}
%\end{split}
%\end{equation}

 We now expand the reduced density matrix in a basis of harmonic oscillator eigenstates ($H_0|i\rangle = \epsilon_0(i) |i\rangle$, $H_1|i'\rangle = \epsilon_1(i') |i'\rangle$,
\begin{equation}\label{eq39}
\begin{split}
    \frac{d\rho_0(i,j)}{dt} = & -\frac{i}{\hbar}(\epsilon_0(i)-\epsilon_0(j))\rho_0(i,j) \\ &
    - \frac{\Gamma}{2\hbar}\sum_{i',k}\tilde{f}(\epsilon_1(i')-\epsilon_0(k))F_{i\rightarrow i'}F_{k\rightarrow i'}\rho_0(k,j) \\ &
    + \frac{\Gamma}{2\hbar}\sum_{i',j'}(1-\tilde{f}(\epsilon_1(j')-\epsilon_0(j)))F_{i\rightarrow i'}F_{j\rightarrow j'}\rho_1(i',j') \\ &
    - \frac{\Gamma}{2\hbar}\sum_{i',k}\rho_0(i,k)\tilde{f}(\epsilon_1(i')-\epsilon_0(k))F_{j\rightarrow i'}F_{k\rightarrow i'} \\ &
    + \frac{\Gamma}{2\hbar}\sum_{i',j'}\rho_1(i',j')(1-\tilde{f}(\epsilon_1(i')-\epsilon_0(i)))F_{i\rightarrow i'}F_{j\rightarrow j'}
\end{split}
\end{equation}
\begin{equation}\label{eq40}
\begin{split}
    \frac{d\rho_1(i',j')}{dt} = & -\frac{i}{\hbar}(\epsilon_1(i')-\epsilon_1(j'))\rho_1(i',j') \\ &
    - \frac{\Gamma}{2\hbar}\sum_{i,k'}(1-\tilde{f}(\epsilon_1(k')-\epsilon_0(i)))F_{i\rightarrow i'}F_{i\rightarrow k'}\rho_1(k',j') \\ &
    + \frac{\Gamma}{2\hbar}\sum_{i,j}\tilde{f}(\epsilon_1(j')-\epsilon_0(j))F_{i\rightarrow i'}F_{j\rightarrow j'}\rho_0(i,j) \\ &
    - \frac{\Gamma}{2\hbar}\sum_{i,k'}\rho_1(i',k')(1-\tilde{f}(\epsilon_1(k')-\epsilon_0(i)))F_{i\rightarrow j'}F_{i\rightarrow k'} \\ &
    + \frac{\Gamma}{2\hbar}\sum_{i,j}\rho_0(i,j)\tilde{f}(\epsilon_1(i')-\epsilon_0(i))F_{i\rightarrow i'}F_{j\rightarrow j'}
\end{split}
\end{equation}
The above equations are referred to as our Floquet Quantum Master equation. Here $\epsilon_0(i)=\hbar\omega(i+\frac{1}{2})$, and $\epsilon_1(i')=\hbar\omega(i'+\frac{1}{2})+\overline{E}_d$. $\overline{E}_d$ is the renormalized impurity energy level.
\begin{equation}\label{eq41}
\begin{split}
    \overline{E}_d \equiv E_d - E_r
\end{split}
\end{equation}
where $E_r\equiv g^2/\hbar\omega$ is the reorganization energy. 
$F$ is the Frank-Condon factor.
\begin{equation}\label{eq42}
\begin{split}
    F_{i\rightarrow i'} = \braket{i'|i} = \int dx\phi_i(x+\sqrt{2}\lambda)\phi_i(x), \lambda\equiv g/\hbar\omega
\end{split}
\end{equation}
where $\phi_i(x)$ is the $i$th eigenfunction of the harmonic oscillator. The Frank-Condon factor can be expressed as \begin{equation}\label{eq43}
\begin{split}
    F_{i\rightarrow i'} = (p!/Q!)^{1/2}\lambda^{Q-p}e^{-\lambda^2/2}L_p^{Q-p}(\lambda^2)[sgn(i'-i)]^{i-i'}
\end{split}
\end{equation}
p(Q) is the minimum (maximum) of $i$ and $i'$, 
and $L_n^m$ is generalized Laguerre polynomial. 
$\tilde{f}(x)$ in the above equations is the modified Fermi function with Floquet replicas, which is given by 
\begin{equation}\label{eq44}
\begin{split}
    \tilde{f}(x) = \sum_{n,m}(i)^n(-i)^m J_n(z) J_m(z) e^{i(n-m)\Omega t} f(x-m\Omega)
\end{split}
\end{equation}
Here we have defined $z=\frac{A}{\hbar\Omega}$. Notice that  $\tilde{f}(x)$ is time dependent. When taking time average over a cycle, we arrive at time-independent modified Fermi function 
\begin{equation}\label{eq45}
\begin{split}
    \overline{\tilde{f}}(x) = \sum_{n} J_n(z)^2 f(x-n\Omega)
\end{split}
\end{equation}
Finally, $\Gamma$ in Eqs. \ref{eq39} and \ref{eq40} is the hybridization function,
\begin{equation}\label{eq46}
\begin{split}
    \Gamma(\epsilon) = 2\pi\sum_k|V_k|^2\delta(\epsilon_k-\epsilon)
\end{split}
\end{equation}
In the wide band limit, we assume that $\Gamma$ is a constant (i.e., does not change with $\epsilon$ or $x$)

\subsection{Floquet Classical Master Equation}

In the limit of $kT > \hbar\omega$, we can treat the nuclear motion classically, such that we will arrive at the Floquet classical master equation (FCME). The FCME is obtained by taking Wigner transform of FQME (Eqs. \ref{eq39} and \ref{eq40}), similar to the steps taken in Ref. \cite{SH_2}: 
\begin{equation}\label{eq47}
\begin{split}
    \frac{\partial P_0(x,p)}{\partial t} = & \frac{\partial H_0(x,p)}{\partial x}\frac{\partial P_0(x,p)}{\partial p} - \frac{\partial H_0(x,p)}{\partial p}\frac{\partial P_0(x,p)}{\partial x} \\ & - \gamma_{0\rightarrow1}P_0(x,p) + \gamma_{1\rightarrow0}P_1(x,p)
\end{split}
\end{equation}
\begin{equation}\label{eq48}
\begin{split}
    \frac{\partial P_1(x,p)}{\partial t} = & \frac{\partial H_1(x,p)}{\partial x}\frac{\partial P_1(x,p)}{\partial p} - \frac{\partial H_1(x,p)}{\partial p}\frac{\partial P_1(x,p)}{\partial x} \\ & + \gamma_{0\rightarrow1}P_0(x,p) - \gamma_{1\rightarrow0}P_1(x,p)
\end{split}
\end{equation}
where
\begin{equation}\label{eq49}
\begin{split}
    \gamma_{0\rightarrow1} = \frac{\Gamma}{\hbar}\tilde{f}(\Delta V)
\end{split}
\end{equation}
\begin{equation}\label{eq50}
\begin{split}
    \gamma_{1\rightarrow0} = \frac{\Gamma}{\hbar}(1-\tilde{f}(\Delta V))
\end{split}
\end{equation}
\begin{equation}\label{eq51}
\begin{split}
    \Delta V = H_1 - H_0 = E_d + \sqrt{\frac{2m\omega}{\hbar}}gx
\end{split}
\end{equation}
$\tilde{f}(\Delta V)$ is given in Eq. \ref{eq44}. The Floquet CME is similar to the non-Floquet CME \cite{SH_2}, except that the Fermi function is modified with Floquet replicas. In the non-Floquet CME, $\gamma_{0\rightarrow1}$ ( $\gamma_{1\rightarrow0}$) is a real number, and can be interpret as the hopping rates. Such that the detailed balance is achieved. In the Floquet CME, however,  the modified Fermi function  $\tilde{f}(\Delta V)$ is time-dependent and complex valued, such that $\gamma_{0\rightarrow1}$ ( $\gamma_{1\rightarrow0}$) cannot be simply interpreted as hopping rates in trajectory based algorithm (see below). Still, in the limit of fast driving (large $\Omega$), we can take the time average of the modified Fermi function, such that $\gamma_{0\rightarrow1}$ ( $\gamma_{1\rightarrow0}$) will become real valued and can be interpreted as hopping rate. Below, we will use trajectories based surface hopping algorithm to solve the Floquet CME. 

\subsection{Floquet Surface Hopping}
As stated above, the Floquet CME is similar to non-Flouqet CME in form, except the hopping rates are complexed valued. We now introduce 3 surface hopping algorithm to solve Floquet CME:
\begin{enumerate}
   \item FSH: in FSH, we interpret the real positive part of $\gamma_{0\rightarrow1}$ as our hopping rates and determine a hopping event from state $\ket{0}$ to state $\ket{1}$ based on $\xi < \Re(\gamma_{0\rightarrow1}) dt$, where $\xi$ is a random number between $0$ and $1$. Similarly, a hopping event from  state $\ket{1}$ to state $\ket{0}$ is determined by $\xi < \Re(\gamma_{1\rightarrow0}) dt$
   \item FaSH: in FaSH, we take the time average of $\tilde{f}(\Delta V)$ over a cycle that becomes $\overline{\tilde{f}}(\Delta V)$, such that the hopping rates are real valued and determine the hopping probability. 
   \item FaSH-density: in FaSH-density, we use the time-averaged $\overline{\tilde{f}}(\Delta V)$ as our hopping rates to propagate nuclear dynamics as same as FasH. In addition, we propagate electron density using $\dot{P_0}= - \gamma_{0\rightarrow1}P_0 + \gamma_{1\rightarrow0}P_1$ and 
$\dot{P_1}= \gamma_{0\rightarrow1}P_0(x,p) - \gamma_{1\rightarrow0}P_1(x,p)$. Here $\gamma_{0\rightarrow1}$ and $\gamma_{1\rightarrow0}$ are non time-averaged, such that we can capture the oscillation in a cycle for electronic dynamics.  
\end{enumerate}

We initialize our nuclei in one well (the unoccupied electronic state) with a Boltzmann distribution. Notice that the initial nuclei temperature can be different from electronic temperature, such that we can simulate energy relaxation of the nuclear kinetic energy.

\section{Results}
We now benchmark our trajectory based algorithms against FQME for electronic population as well as nuclear kinetic energy. 

\begin{figure*}[htbp]
\centering
\includegraphics[scale=0.12]{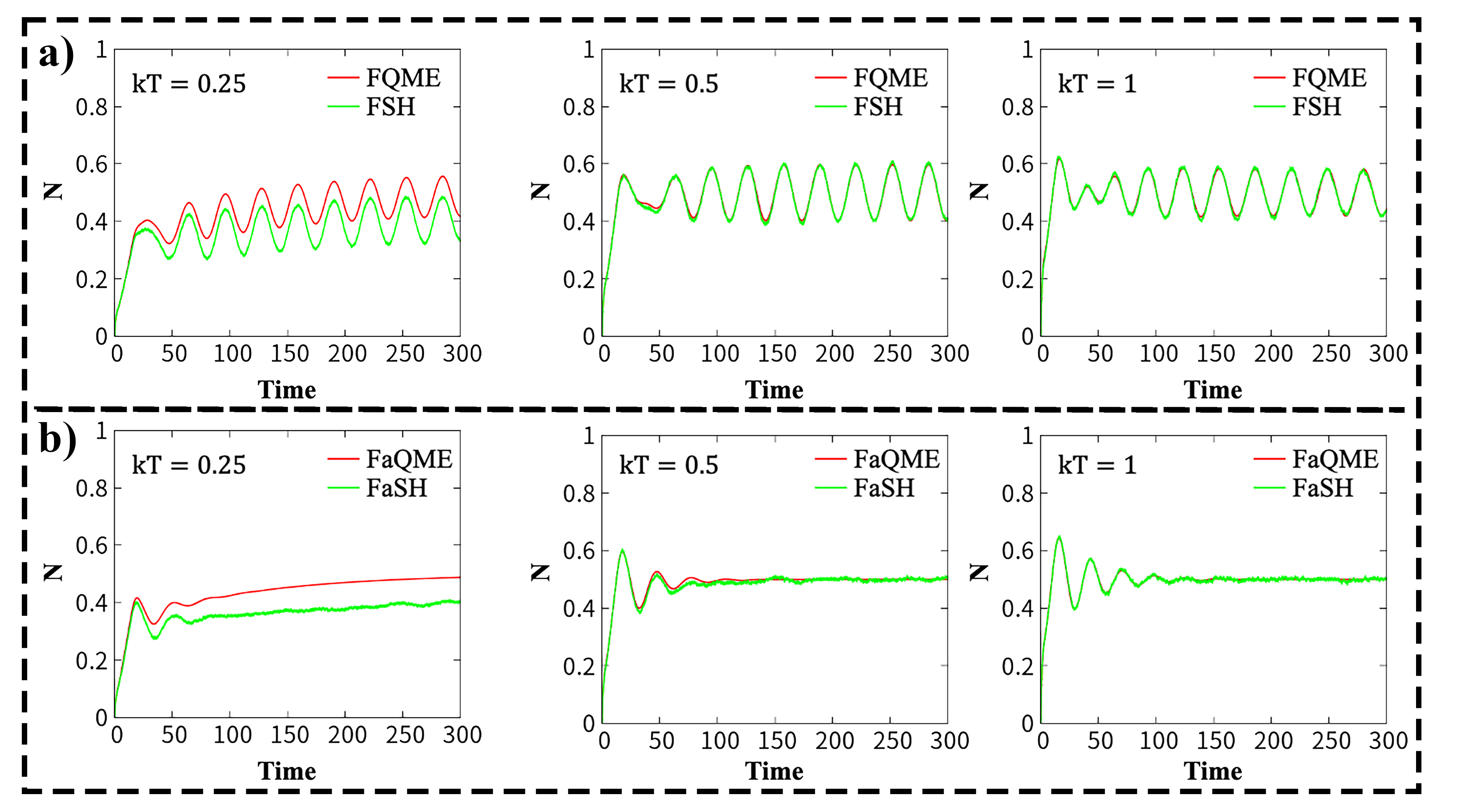}
\caption{The impurity electronic population as function of time for a) non-time-averaged FQME and FSH; b) time averaged FaQME and FaSH.
$\Gamma = 1$, $\hbar\omega=0.3$, el-ph coupling $g=0.75$, 
$\overline{E}_d=0$. The driving amplitude is $A=0.2$ and the driving frequency is $\Omega=0.2$. Note that FSH and FaSH agree with FQME and FaQME at high temperature. At time zero, the phonon is prepared to be equilibrated thermally at unoccupied impurity level.}
\label{fig1}
\end{figure*}

We first look at the the electronic population with periodic driving as a function of time. In Fig. \ref{fig1}a, we show the electronic population dynamics from FQME and FSH at different temperatures ($kT = 0.25, 0.5, 1$). Here, we set nuclear vibration frequency as $\hbar\omega = 0.3$. As expect, at high temperature ($kT>\hbar\omega$), FSH agrees with FQME very well; whereas at low temperature, FSH shows discrepancy as compared to FQME. Similar trend is observed for FaSH and FaQME as shown in Fig. \ref{fig1}b: in the high temperature limt, FaSH agrees with FaQME very well. Notice that, the time-averaged floquet results (FaSH or FaQME) do not show oscillations and reach to a steady state in the long time. The non time-averaged methods (FSH or FQME) do show the oscillations due to the non vanishing driving. In the long time, FSH or FQME reaches to a cycle limit instead of a stead state. The frequency of the oscillation in electronic population is equal to the frequency of the driving (we have set $\Omega = 0.2$).

\begin{figure*}[htbp]
\centering
\includegraphics[scale=0.15]{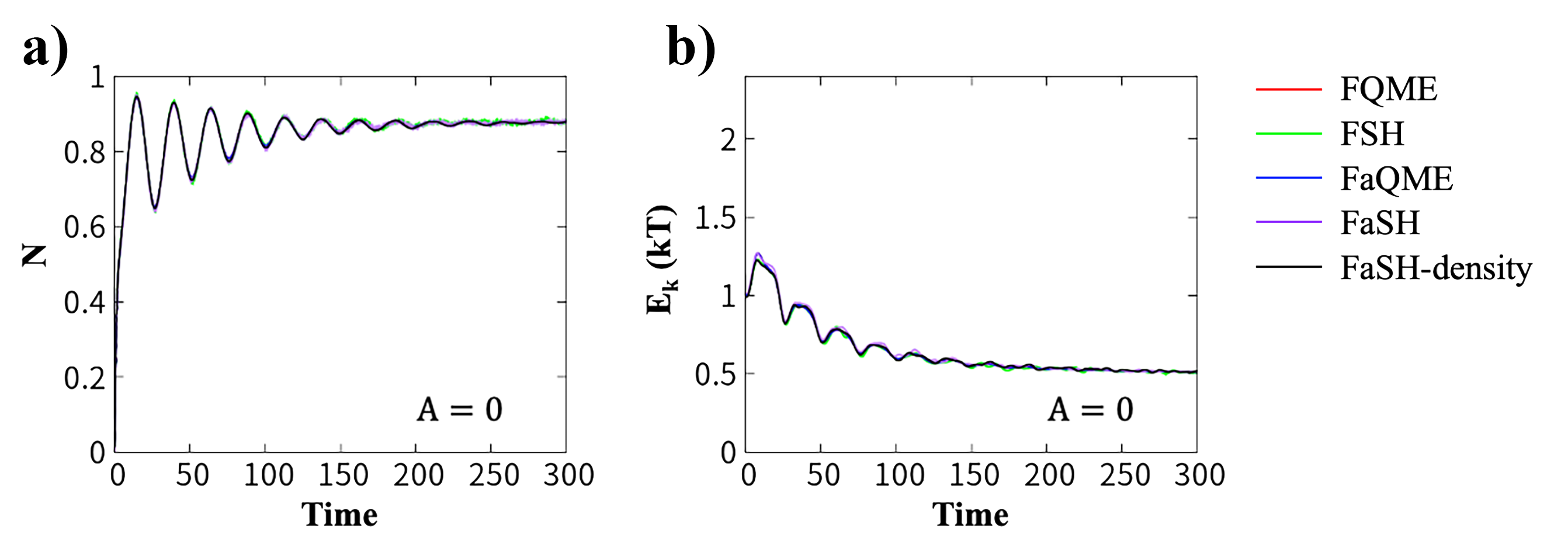}
\caption{We show a) the impurity electronic population and b) the phonon relaxation as a function of time without Floquet engineering for the five methods, i.e., FQME, FSH, FaQME, FaSH, and FaSH-density. 
$kT=1$, $\Gamma = 1$, $\hbar\omega=0.3$, el-ph coupling $g=0.75$, 
$\overline{E}_d=-2$. Note that without Floquet engineerings, the five approaches show exact the same results.}
\label{fig2}
\end{figure*}

To further verify our methods, we show the electronic population and phonon relaxation dynamics 
as a function of time without any Floquet driving in Fig. \ref{fig2}. In such a case, we expect that the five algorithms (FQME, FSH, FaQME, FaSH, and FaSH-density) should all agree with each other at high temperature. Indeed, for the case of $kT > \hbar\omega$ ($kT = 1$, $\hbar\omega = 0.3$), the algorithms all agree with each other for both electronic population (Fig. \ref{fig2}a) as well as nuclear kinetic energy (Fig. \ref{fig2}b). Obviously, without any Floquet driving, both electronic population and nuclear kinetic energy reach to a steady state or equilibrium state. Indeed, the steady state kinetic energy is $E_k = 1/2 kT$ and electronic population is $N = f(\overline{E}_d)$. Here $kT$ is the temperature of the electronic bath. Below, we will mainly look at the high temperature limit where the classical nuclei is valid.

\begin{figure*}[htbp]
\centering
\includegraphics[scale=0.13]{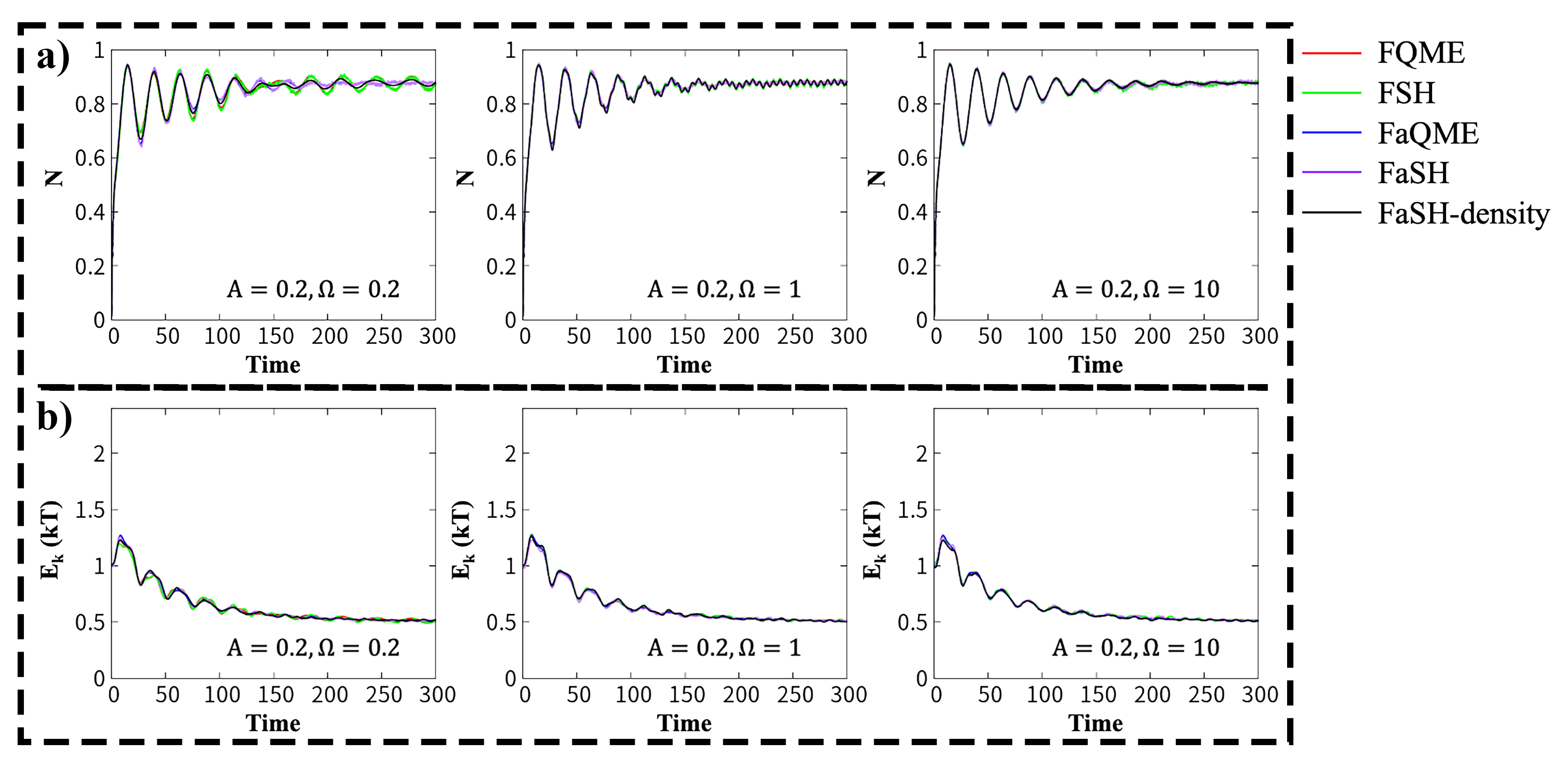}
\caption{a) The impurity electronic population as function of time under 
a small strength of driving amplitude that is comparable with the nuclear oscillator ($A=0.2$)
with three degrees of driving frequencies ($\Omega = 0.2,1,10$).
$kT=1$, $\Gamma = 1$, $\hbar\omega=0.3$, el-ph coupling $g=0.75$, 
$\overline{E}_d=-2$. Five methods, i.e., FQME, FSH, FaQME, FaSH, and FaSH-density are put together for comparison. Note that in small strength of the drivings, five methods give nearly the same feature of electronic and nuclear dynamics with that in Fig. \ref{fig2}, regardless of the driving frequencies.}
\label{fig3}
\end{figure*}

In Fig. \ref{fig3}, we benchmark our algorithms for relatively small driving Amplitude with different driving frequencies. Here, the driving amplitude is $A=0.2$, which is comparable to the nuclear frequency ($\hbar\omega=0.3$). In such a case, the five methods all agree with each other regardless of the driving frequencies. That being said, the non time-averaged methods (FSH, FQME, and FaSH-density) do show small oscillations in electronic dynamics. The steady state in this case is very close to the equilibrium state with $E_k = 1/2 kT$ and $N = f(\overline{E}_d)$.

\begin{figure*}[htbp]
\centering
\includegraphics[scale=0.13]{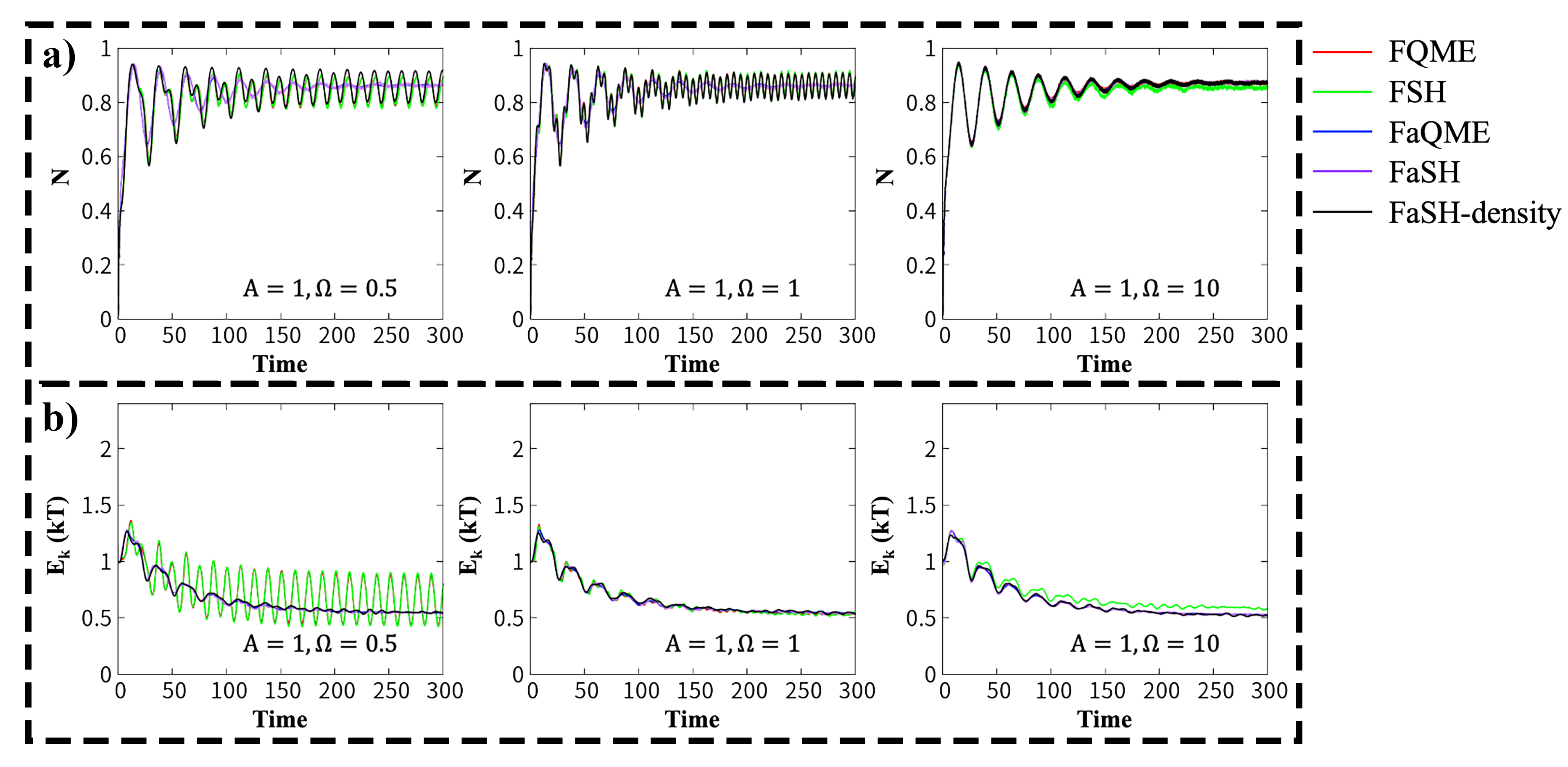}
\caption{a) The impurity electronic population as function of time under 
a medium strong driving amplitude ($A=1$)
with three degrees of driving frequencies ($\Omega = 0.5,1,10$).
$kT=1$, $\Gamma = 1$, $\hbar\omega=0.3$, el-ph coupling $g=0.75$, 
$\overline{E}_d=-2$. We compare five methods, i.e., FQME, FSH, FaQME, FaSH, and FaSH-density for intermediate driving amplitude. Note that the time-averaged FaQME and FaSH cannot capture the oscillation feature both in electronic and nuclear dynamics, especially under slow and medium quick drivings ($\Omega=0.5$ or $\Omega=1$). FSH does not obey detailed balance, resulting in higher effective tempeature at large driving frequency ($\Omega=10$). FaSH-density method is enable to capture the oscillation feature in electronic dynamics and results in correct effective temperature at large $\Omega$. }
\label{fig4}
\end{figure*}

We now turn to the intermediate driving amplitude, where we have set $A = 1$. In such a case, the five methods start to show deviations as shown in Fig. \ref{fig4}. For smaller driving frequency (e.g. $\Omega = 0.5$), the oscillation feature from FQME and FSH becomes stronger both in electronic populaiton and nuclear dynamics, whereas FaSH and FaQME miss the feature completely. FaSH-density shows oscillation in electronic dynamics and fail to reproduce the feature in nuclear dynamics. For slightly larger driving frequency ($\Omega = 1$ or $10$), the oscillation feature becomes weaker. However, FSH starts to show deviations in the steady state at very large driving frequency ($\Omega = 10$), where the kinetic energy is higher than the results from FQME (and others). Consequently, the electronic population from FSH is smaller than the results from other methods. This deviation under quick drivings is mainly derived from disregarding the negative part of the hopping rate in the FSH algorithm. This violates the detailed balance. As a result, the effective temperature of the system from FSH is higher than the true effective temperature from other methods. Note also that, the true effective temperature is slightly different from equilibrium temperature. This is due to the effects of the driving.

\begin{figure*}[htbp]
\centering
\includegraphics[scale=0.13]{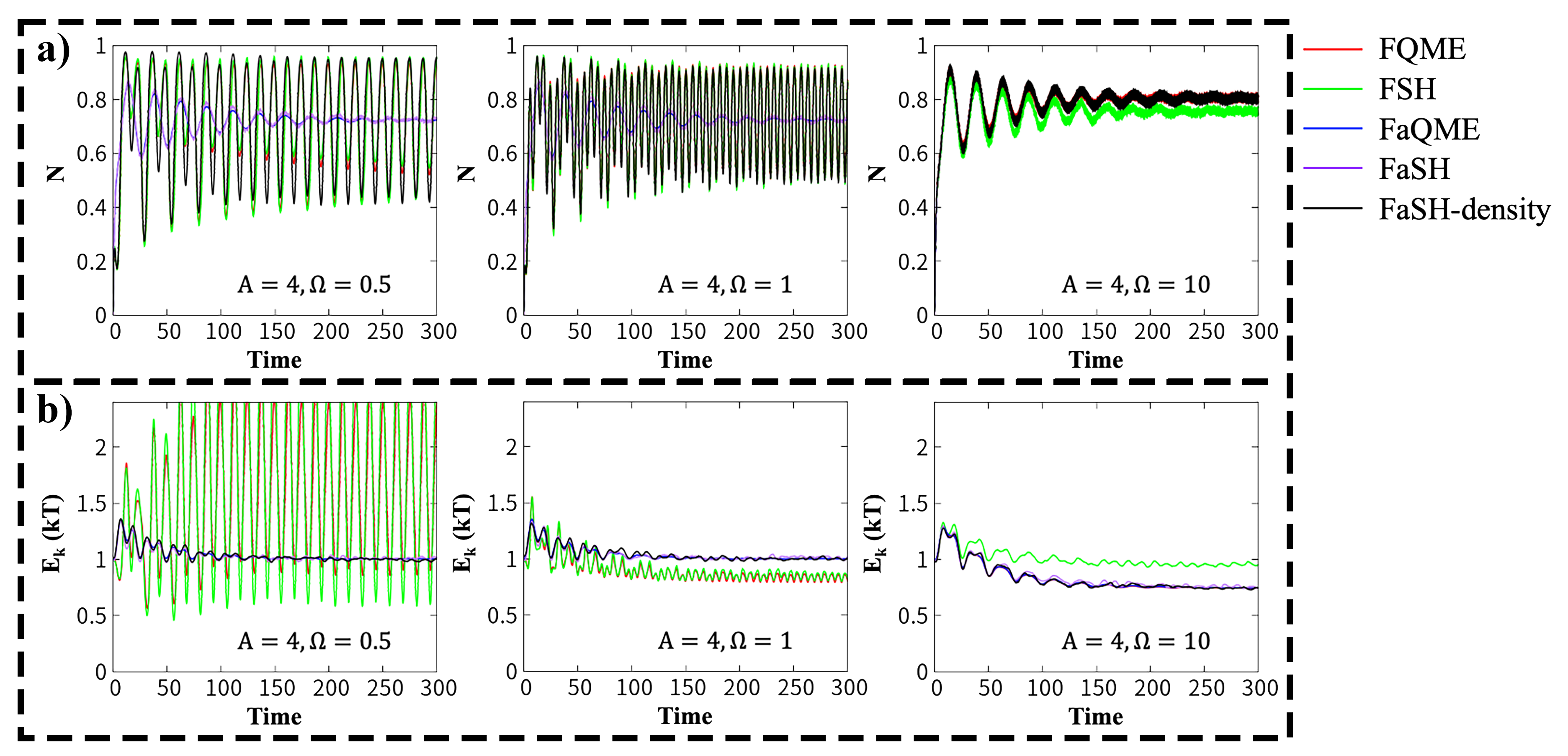}
\caption{a) The impurity electronic population as function of time under a strong driving amplitude ($A=4$)
for small and large driving frequencies ($\Omega = 0.5,1,10$).
$kT=1$, $\Gamma = 1$, $\hbar\omega=0.3$, el-ph coupling $g=0.75$, 
$\overline{E}_d=-2$. Again, the time-averaged FaQME and FaSH cannot capture the oscillation feature under slow and medium quick drives ($\Omega=0.5$ or $\Omega=1$). FSH gives rise to wrong steady-state results under quick drive ($\Omega=10$). FaSH-density performs the best, which is enable to capture the oscillation feature for electronic dynamics and give rise to proper steady-state results.}
\label{fig5}
\end{figure*}

Finally, we show the case of very strong driving amplitude ($A=4$) in Fig. \ref{fig5}. In such a case, the oscillation feature becomes very strong, particularly for small driving amplitude. Again, FSH works very well in small driving frequency, but fails to produce the steady state electronic population and nuclear kinetic energy at large driving frequency. FaSH misses all oscillation feature at small driving frequency, but does reach to correct steady state. FaSH-density performs the best, working both in small driving frequency and large driving frequency.

\section{Conclusions}
In summary, we derived the Floquet quantum master equation to study the dynamics of a periodically driven system near a metal surface. In the high temperature limit, the FQME reduces to a FCME. We proposed three surface hopping algorithms (FSH, FaSH, and FaSH-density) to solve the FCME. In the limit of small driving frequency, FSH works very well, which captures the full dynamics for both electronic population and nuclear kinetic energy as benchmark against FQME. In the large driving frequency, FSH fails to produce the correct steady states. This is due to the fact that the hopping rates are complex valued and we throw out the negative hopping rates in FSH. The FaSH works well in the large driving frequency, but fails to reproduce the oscillation feature at small driving frequency. The FaSH-density performs the best, which reproduces the full electronic dynamics at small and large driving frequencies and reaches to the correct steady states for nuclear motion. We expect that our algorithms will be very useful to study non-adiabatic dynamics with periodic driving in open quantum systems. 

\begin{acknowledgments}
We thank Amikam Levy, Jacob B\"{a}tge, and Michael Thoss for useful and inspiring conversations. This material is based upon work supported by National Science Foundation of China (NSFC No. 22273075)
\end{acknowledgments}

\bibliography{reference}
\end{document}